\documentclass[aps, prx,superscriptaddress, twocolumn,secnumarabic,nobalancelastpage,nofootinbib]{revtex4-1} 
\usepackage{graphicx}
\usepackage{amsmath,amssymb}
\usepackage[utf8]{inputenc}
\usepackage[T1]{fontenc}
\usepackage{lmodern}
\usepackage{natbib}
\usepackage{hyperref}
\usepackage{xcolor}
\hypersetup{colorlinks=true,citecolor={blue},linkcolor={blue},urlcolor={blue}}
\usepackage{units}
\usepackage[english]{babel}

\usepackage[normalem]{ulem}
\usepackage{upgreek}
\usepackage{wasysym}
\usepackage{soul}
\makeindex

\begin{document}

\definecolor{orange}{RGB}{255, 69, 0}
\definecolor{green}{RGB}{26,148,49}
\newcommand{\lp}[1]{{\color{green}{#1}}}
\setstcolor{red}
\newcommand{\hs}[1]{\textcolor{purple}{#1}}
\newcommand{\jt}[1]{\textcolor{orange}{#1}}
\newcommand{\Sz}{S_{z}} 
\newcommand{\SzR}{S_{z_{R}}} 
\newcommand{\SzL}{S_{z_{L}}} 

\title{Polariton spin jets through optical control}

\date{\today}

\author{L. Pickup}
\affiliation{School of Physics and Astronomy, University of Southampton, Southampton, SO171BJ, United Kingdom}

\author{J. D. T\"opfer}
\affiliation{School of Physics and Astronomy, University of Southampton, Southampton, SO171BJ, United Kingdom}

\author{H. Sigurdsson}
\affiliation{School of Physics and Astronomy, University of Southampton, Southampton, SO171BJ, United Kingdom}
\affiliation{Skolkovo Institute of Science and Technology Novaya St., 100, Skolkovo 143025, Russian Federation}

\author{P. G. Lagoudakis}
\email{Pavlos.Lagoudakis@soton.ac.uk}
\affiliation{School of Physics and Astronomy, University of Southampton, Southampton, SO171BJ, United Kingdom}
\affiliation{Skolkovo Institute of Science and Technology Novaya St., 100, Skolkovo 143025, Russian Federation}

\begin{abstract}
We demonstrate spin polarized jets in extended systems of ballistic exciton-polariton condensates in semiconductor microcavities using optical non-resonant excitation geometries. The structure of the spin jets is determined by the digitally reprogrammable, spatially non-uniform, degree of circular polarization of the excitation laser. The presence of the laser excitation, strong particle interactions, and spin-relaxation leads to a tunable spin-dependent potential landscape for polaritons, with the appearance of intricate polarization patterns due to coherent matter-wave interference. Our work realizes polarization-structured coherent light sources in the absence of gauge fields.
\end{abstract}

\maketitle

\section{Introduction}
The spin degree of freedom in spinor condensed matter systems provides a parameter in which information can be encoded and, as such, has retained interest over a number of years and platforms~\cite{malajovich_persistent_2001,bhatti_spintronics_2017} for spintronic applications. Controllable generation of a spin current, like the persistent spin helix effect~\cite{Koralek_Nature2009}, is one of the desirable features in spin-dependent devices with studies being pushed beyond electronic systems~\cite{Han_NatMat2020}. 
%
%
In semiconductor microcavities, exciton-polaritons (hereafter polaritons) are characterized by high temperature condensation~\cite{kasprzak_boseeinstein_2006, Carusotto_RMP2013}, strong nonlinearity, ultrafast spin dynamics \cite{lagoudakis_stimulated_2002}, and a multitude of optical-based techniques to manipulate their spin state. They offer a promising platform to investigate spin dynamics in extreme condensed matter settings~\cite{Shelykh_PRB2004, askitopoulos_nonresonant_2016, ohadi_spin_2017, pickup_optical_2018, askitopoulos_all-optical_2018, Sich_ACSPho2018, klaas_nonresonant_2019, del_valle-inclan_redondo_observation_2019}, and for spinoptronic applications~\cite{Liew_PRL2008, Shelykh_PRB2009, Amo_NatPho2010, cerna_ultrafast_2013, dreismann_sub-femtojoule_2016, Sedov_LScApp2019, Mandal_PRApp2019}.

Polaritons are the bosonic quasi-particles formed in the strong coupling regime through the hybridization of light confined in a Fabry-P\'{e}rot microcavity and electronic transitions in embedded intra-cavity quantum wells. The resulting polariton modes are part light, part matter particles with two possible projections of their spin ($\pm1$) along the growth axis of the cavity. The optical selection rules enable direct measurement of the spinor polariton structure via standard polarization resolved imaging of photoluminescence (PL). Conversely, one can excite directly polariton spin states using polarized resonant lasers~\cite{Paraiso_NatMat2010, cerna_ultrafast_2013}, or indirectly using polarized non-resonant lasers~\cite{askitopoulos_nonresonant_2016, klaas_nonresonant_2019, gnusov2020optical}.

Spinoptronic application of polariton condensates involves control of their nonlinearity through the condensate density and of real, or effective, gauge fields acting on the polariton spin, like the optical spin Hall effect~\cite{Leyder_NatPhy2007, Kamman_PRL2012, Cilibrizzi_PRB2015, Sich_ACSPho2018}. Control over the condensate density in each spin component is realized through the excitation intensity and polarization~\cite{Anton_PRB2015, ohadi_spin_2017, pickup_optical_2018, del_valle-inclan_redondo_observation_2019, askitopoulos_nonresonant_2016, askitopoulos_all-optical_2018, Ryzhov_PRR2020, gnusov2020optical}, whereas a more deterministic control over the spin dynamics can be employed using external magnetic fields~\cite{Walker_PRL2011, kulakovskii_magnetic_2012, Pietka_PRB2015, caputo_magnetic_2019}, electric fields~\cite{dreismann_sub-femtojoule_2016}, or a combination of both~\cite{Lim_NatComm2017}.


The system's dependencies on the polarization of the excitation laser under various spin-coupling circumstances has been extensively studied for polariton condensates in optical (laser induced) traps~\cite{ohadi_spin_2017, pickup_optical_2018, del_valle-inclan_redondo_observation_2019, askitopoulos_nonresonant_2016, askitopoulos_all-optical_2018, Ryzhov_PRR2020, gnusov2020optical} and in fabricated patterned photonic lattices~\cite{Klemb_Nature2018, klaas_nonresonant_2019, Jamadi_LScApp2020}. However, extended systems of highly energetic ballistically expanding polariton condensates~\cite{Ohadi_PRX2016, topfer_time-delay_2020, Pickup_NatComm2020}, which possess their own unique physical properties~\cite{topfer_time-delay_2020}, have not been properly explored in terms of spin. Not only are ballistic condensates characterized by strong particle currents which can couple them together over long distances but they also qualify for large coherent lattices~\cite{topfer2020engineering}. Systems of such condensates might therefore present new ways of implementing strongly spin-polarized particle currents with long range coherence.


In this report, we demonstrate the appearance and control of strongly polarized spin jets (currents) in an extended system of two coupled polariton condensates. The spin jets originate from the strongly interactive nature of polaritons with their non-condensed particle background. This interaction results in a local repulsive potential gradient which ejects polaritons away from the pump spots resulting in strong interference between neighboring condensates. The pattern of the spin jets is found to depend critically on the degree of circular polarization (DCP) of the non-resonant pumps. Our results pave the way towards controllable structures of spin polarized matter wave fluids.


\begin{figure}[!t]
	\centering
	\includegraphics[width=8.6cm]{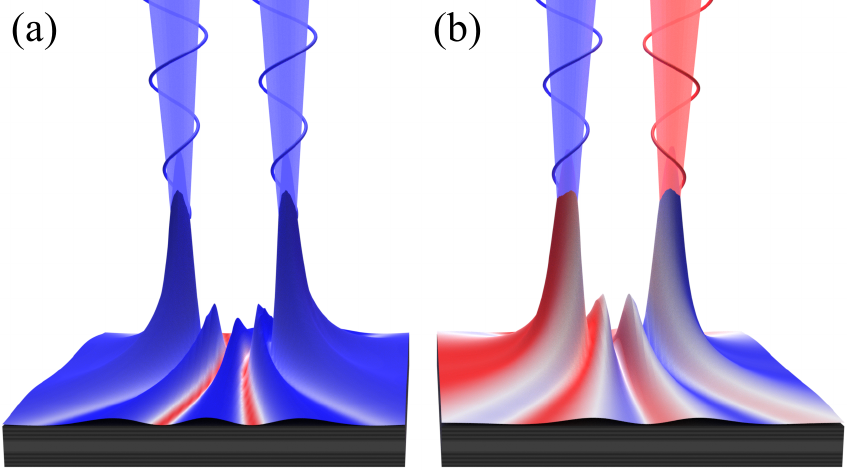}
	\caption{Schematic representing two pump polarization geometries where (a) the pumps are co-circularly polarized ($\SzL=\SzR$) and (b) the pumps are cross-circularly polarized  ($\SzL=-\SzR$). The height of the surface represents the spatial polariton condensate density from simulation. The red-white-blue colormap shows the degree of circular polarisation [see Eq.~\eqref{eq.Szc}] with red representing $\Sz=-1$ (spin-down polaritons) and blue $\Sz=1$ (spin-up polaritons).}
	\label{Fig_Schematic}
\end{figure}

\section{Results}
The sample used in our experiment is a planar $2\lambda$ GaAs microcavity with eight 6 nm InGaAs intra-cavity QWs configured in pairs around the three anti-nodal positions of the cavity mode with an additional QW at the outermost node either side of the cavity~\cite{Cilibrizzi_APL2014}. The sample is cooled to $\sim6$ K in a cold finger flow cryostat and is excited by a continuous wave laser blue-detuned to a reflectivity minimum above the cavity stop band ($\lambda \approx 800\;\mathrm{\mu m})$ to maximize coupling efficiency of light into the system without directly exciting the polariton modes. The excitation laser is spatially modulated into two approximately Gaussian spots (full-width-half-maximum $\approx 2\;\mathrm{\mu m}$) separated by a distance `$d$' using a reflective phase-only liquid crystal spatial light modulator (SLM). To attain arbitrary spatial control of the pump polarization the spatially modulated beam is focused onto a secondary translucent liquid crystal SLM before being projected onto the sample surface.

\subsection{Co- and cross-polarized pumps}
\begin{figure}[!t]
	\centering
	\includegraphics[width=8.6cm]{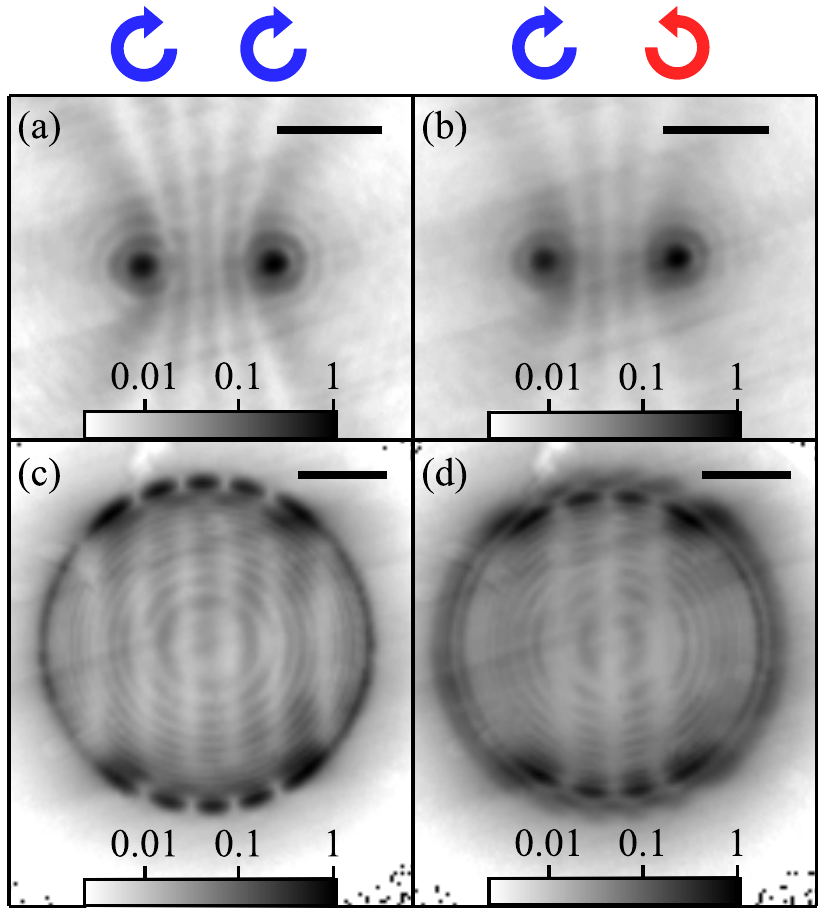}
	\caption{Characterization of a spinor polariton dyad (two condensate system) with a pump separation distance of approximately $12.5\;\mathrm{\mu m}$. (a,b) and (c,d) Experimental real-space and momentum-space PL intensity distributions respectively for co-circularly (a,c) and cross-circularly (b, d) polarized pumps. The two high intensity (dark colored) spots in real space represent the individual condensate centers excited by their respective pumps. The solid black bars at the top right hand corner or each panel are scale bars representing $10$ $\mu$m in (a,b) and $1$ $\mu$m$^{-1}$ in (c \& d).}
	\label{Fig1}
\end{figure}

We consider two different experiments as shown schematically in Fig.~\ref{Fig_Schematic}(a \& b). In the former experiment the two highly-polarized pump beams have the same DCP, $\SzL=\SzR \approx 0.84$, where $\SzL$ and $\SzR$ correspond to the left ($L$) and right positioned ($R$) pump beams respectively in the cavity plane. In the latter experiment the pumps are anti-circularly polarized $\SzL = -\SzR \approx 0.84$. We note that the use of $|\Sz| \approx 0.84$ instead of higher absolute value of DCP is due to technical limitations of the efficiency of the transmissive SLM at the excitation wavelength. Figure~\ref{Fig1} compares the two experiments for a condensate separation distance of $d \approx 12.5\; \mathrm{\mu m}$, where Figs.~\ref{Fig1}(a) and~\ref{Fig1}(c) correspond to $\SzL=\SzR=0.84\pm0.01$ and Figs.~\ref{Fig1}(b) and~\ref{Fig1}(d) correspond to $\SzL=+0.84\pm0.01$ and $\SzR=-0.84\pm0.01$. Figures~\ref{Fig1}(a,b) and~\ref{Fig1}(c,d) show the experimentally measured real space and momentum space PL intensity distributions of the interacting polariton condensates respectively (not polarization resolved). The position of the Gaussian pump beams coincides with the two strong emission intensity peaks shown in the real space PL. The presence of clear interference fringes in the time averaged images Figs.~\ref{Fig1}(a-d) is the result of robust phase-locking between the two ballstically expanding condensate nodes~\cite{topfer_time-delay_2020}. 

For co-circularly polarized pumps where $\SzR = \SzL = 0.84\pm0.01$ we observe that the system condenses into a state characterized by the three interference fringes between the condensate centers in Fig.~\ref{Fig1}(a). Keeping the separation distance $d$ constant and changing the DCP of the right pump to have $\SzR = -0.84\pm0.01$ , while retaining $\SzL = 0.84\pm0.01$, we observe that the system now dominantly condenses into a state with two fringes between the condensate centers (see Fig.~\ref{Fig1}(b)). 

To understand why the condensate interference pattern changes when switching the sign of the DCP for one of the pumps we write the following single-particle two-dimensional (2D) Schr\"{o}dinger equation for the spinor polaritons in the circular polarization basis of the photons which corresponds to the spin-up and spin-down basis of the polaritons,
\begin{equation}
    i \hbar \frac{\partial \psi_\pm}{\partial t} = \left[ - \frac{\hbar^2 \nabla^2}{2m^*} + V_\pm(\mathbf{r}) - \frac{i \hbar \gamma}{2} \right] \psi_\pm.
\end{equation}
Here, $m^*$ is the polariton mass, $\gamma$ is the polariton decay rate, and $V_\pm$ is the potential of each spin component. The potential $V_\pm$ is directly proportional to the background ensemble of uncondensed particles which is referred to as the ``exciton reservoir`` density $n_\pm$~\cite{Wertz_NatPhys2010, Anton_PRB2013}
\begin{equation} \label{eq.pot0}
    V_\pm(\mathbf{r}) = \hbar \left(g_1 + i\frac{R}{2} \right) n_\pm + \hbar g_2 n_\mp.
\end{equation}
Here $g_{1,2}$ are the anisotropic co- and cross-spin interaction strengths of polaritons with the excitonic reservoir and $R>0$ is the scattering rate of polaritons into the condensate. The reservoir of excitons is in-turn proportional to the laser power density,
\begin{equation} \label{eq.res}
    n_\pm \propto (1 + \eta)P_\pm(\mathbf{r}) + \eta P_\mp(\mathbf{r})\hs{.}
\end{equation}
Here $P_\pm$ are the spin-up and spin-down intensity components of the pump laser respectively. The parameter $\eta$ quantifies the contribution of exciton spin-relaxation within the reservoir which mixes the exciton spins.

\begin{figure}[t]
	\centering
	\includegraphics[width=8.6cm]{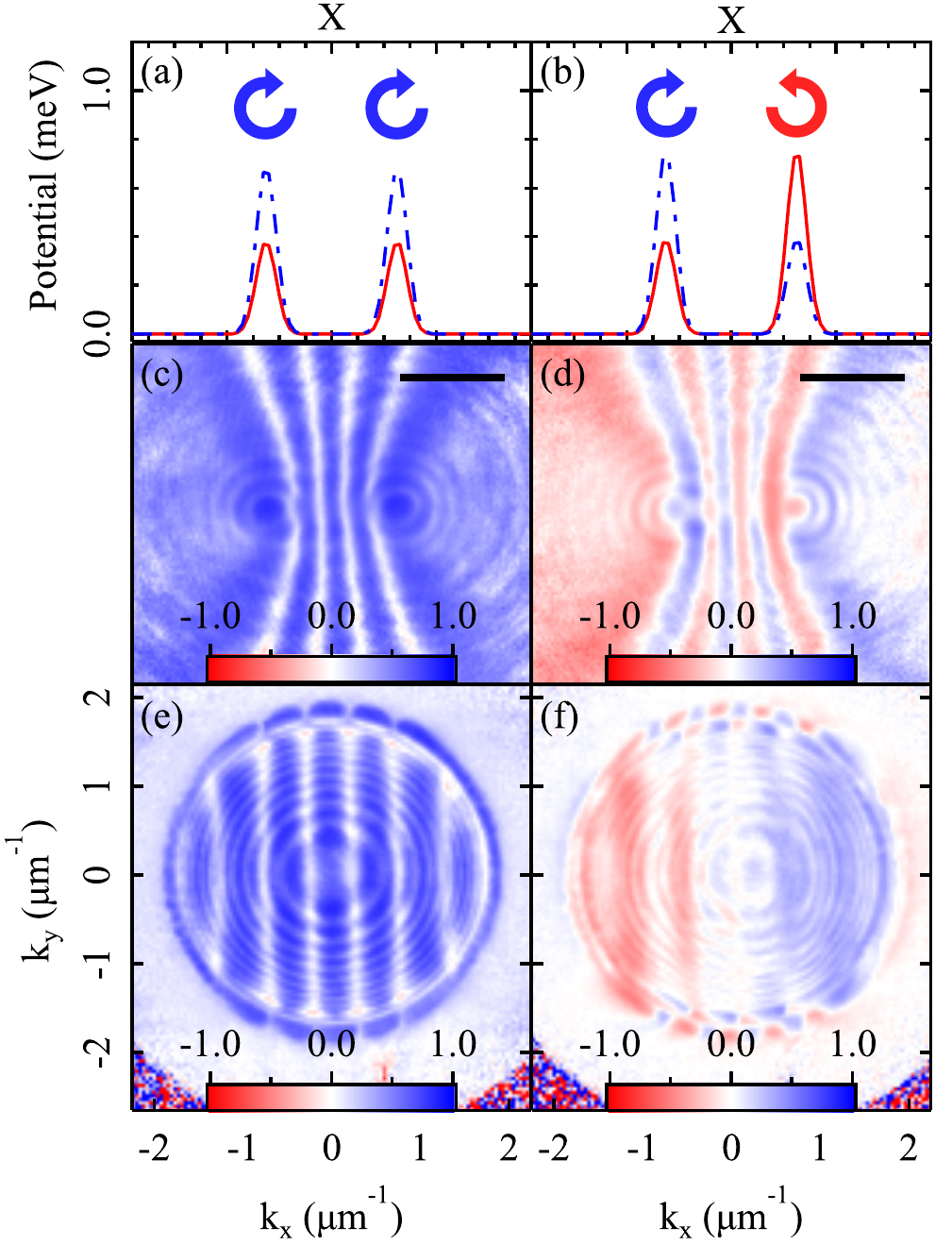}
	\caption{DCP maps for a dyad separation distance $d=12.5~\mu$m for co- and cross-circularly polarized pump beams. (a \& b) Real parts of the simulated potential along the $x$-axis connecting the condensates felt by spin-up ($V_+$, blue dashed lines) and spin-down ($V_-$, red solid lines) polaritons for the two pump polarization geometries shown graphically by the circular arrows. (c \& d) Experimentally measured real space $\Sz^{(c)}$ maps for parallel and cross circularly polarized pumps respectively. (e \& f) Experimentally measured momentum space $\Sz^{(c)}$ maps for parallel and cross circularly polarized pumps respectively. The solid black bars in (c \& d) represent $10~\mu$m.}
	\label{Fig2}
\end{figure}

From the above equations it becomes clear that if the two pump beams are co-circularly polarized like in Figs.~\ref{Fig1}(a) and~\ref{Fig1}(c) then one has 
\begin{equation} \label{eq.pot1}
    V_\pm(\mathbf{r}) = V_\pm(-\mathbf{r}),
\end{equation}
where $V_+(\mathbf{r}) \propto V_-(\mathbf{r})$. That is, the potential landscapes experienced by each polariton spin component are symmetric about the origin of the dyad (its ``center of mass'') and the same up to a scalar factor (see Fig.~\ref{Fig2}(a)). On the other hand, if the pump beams are cross-circularly polarized then reflection symmetry in the $x$-coordinate connecting the condensate centers is broken,
\begin{equation} \label{eq.pot2}
    V_\pm(\mathbf{r}) \neq V_\pm(-\mathbf{r}),
\end{equation}
and one has instead $V_+(\mathbf{r}) = V_-(-\mathbf{r})$. The potentials $V_\pm$ are no longer symmetric about the origin but rather are flipped around the $y$-axis with respect to the two spins (see Fig.~\ref{Fig2}(b)). This dramatic change in the shape of the potentials seen by each spin underpins the different interference patterns observed here. Indeed, the difference between Figs.~\ref{Fig1}(a,c) and~\ref{Fig1}(b,d) is a redistribution of potential energy such that the obtained kinetic energy of the polaritons is also redistributed, thus changing the number of interference fringes.

Another striking consequence of the creation of these asymmetric spin potentials are the polarization patterns formed. Figures~\ref{Fig2}(c,d) show the experimental real-space maps of the condensate DCP defined as,
\begin{equation} \label{eq.Szc}
    S_z^{(c)}(\mathbf{r}) = \frac{I_+(\mathbf{r}) - I_-(\mathbf{r})}{I_+(\mathbf{r}) + I_-(\mathbf{r})}, 
\end{equation}
where $I_\pm(\mathbf{r}) \propto \int|\psi_\pm(\mathbf{r},t)|^2\,\mathrm{d}t$ are the right circular polarized (RCP) and left circular polarized (LCP) time integrated cavity PL emission intensities.

In Fig.~\ref{Fig2}(c) the spatial $S_z^{(c)}(\mathbf{r})$ map shows a dominance in the intensity of RCP PL depicted with blue color corresponding to a larger population of spin-up condensate polaritons. This is a result of the pump partially spin-polarizing the background reservoir, and that scattering of polaritons into the condensate is dominantly spin preserving~\cite{Ciuti_PRB1998, Renucci_PRB2005}. This results in a condensate polarization aligned with that of the pump. The corresponding measured momentum-space map of the condensate DCP, $\Sz^{(c)}(\mathbf{k})$, which is defined analogous to Eq.~\ref{eq.Szc}, is shown in Fig.~\ref{Fig2}(e).

Conversely, for the cross-circularly polarized pump regime ($\SzL\approx-\SzR$) the spatial map of $S_z^{(c)}(\mathbf{r})$ demonstrates an inversion with respect to the pump polarization profile. The left and right hand pumps preferentially provide gain to spin-up and spin-down polaritons respectively but the PL polarization shows a dominance of  spin-down polaritons to the left of the dyad and spin-up polaritons to the right. The origin of this flipped spin pattern are counter propagating spin currents as can be seen in the momentum space $\Sz^{(c)}(\mathbf{k})$ map in Fig.~\ref{Fig2}(f). These counter propagating currents result from the asymmetric potentials that each spin is subject to given by Eq.~\eqref{eq.pot2} (see Fig.~\ref{Fig2}(b)) and result in unequal transmission and reflection amplitudes of the polariton spins in each direction along the dyad axis. 

As shown by Eqs.~\eqref{eq.pot0} and~\eqref{eq.res} there are two main contributions to the asymmetric potential which explains the observations in Figs.~\ref{Fig1} and~\ref{Fig2}. They are the reservoir spin relaxation $\eta$, and the attractive cross-spin interactions between condensate polaritons and reservoir excitons $g_2$. It is generally known however that cross-spin interactions are weaker than same-spin interactions, with an estimate of $-0.5 \lesssim g_2/g_1 \leq 0 $~\cite{Ciuti_PRB1998, Renucci_PRB2005, Krizhanovskii_PRB2006}. On the other hand, spin-relaxation of excitons~\cite{Maialle_PRB1993} can be observed as PL depolarization below the condensation threshold and can be considerably fast compared to other timescales of both the reservoir and the condensate~\cite{Ohadi_PRL2012, klaas_nonresonant_2019}. In modelling our system using the 2D spinor Gross-Pitaevskii equation (2DGPE)  [see Appendix~\ref{Methods_Simulation}] we indeed find that the spin relaxation rate is the critical parameter to produce the observed results.

\begin{figure}[t]
	\centering
	\includegraphics[width=8.6cm]{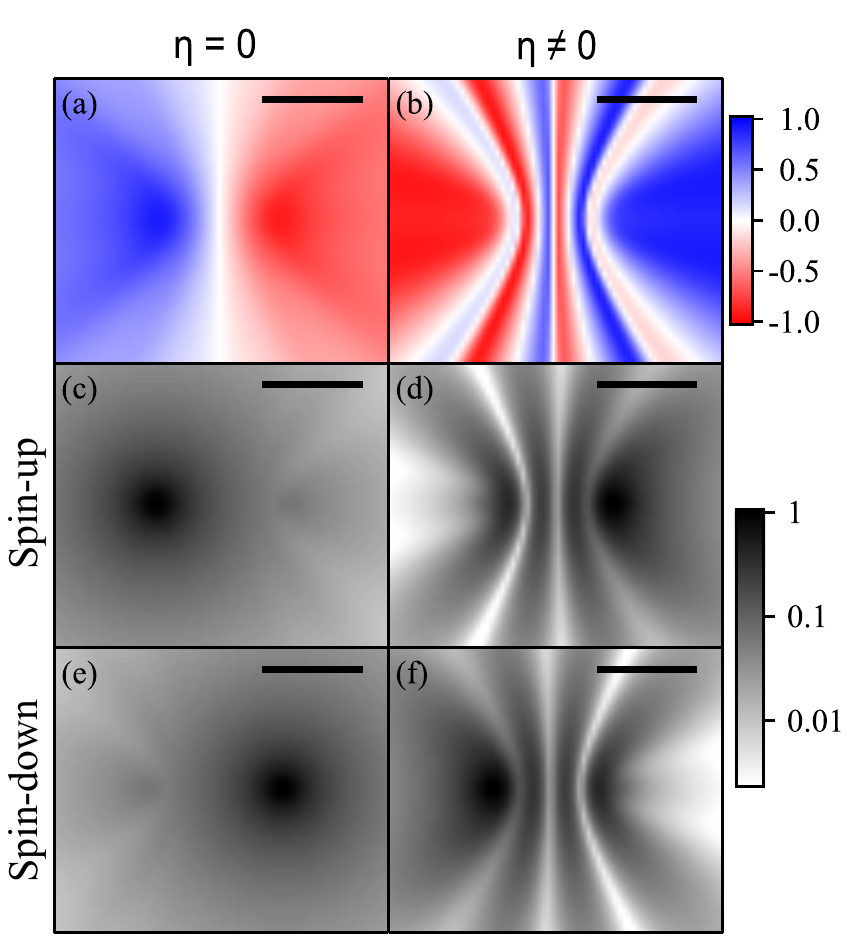}
	\caption{Time integrated solutions from 2DGPE simulations highlighting the effect of accounting for spin relaxation in the reservoirs. For a dyad with a separation distance $d=12.5\;\mathrm{\mu m}$ and $\SzL=0.84$, $\SzR=-0.84$, (a,b) spatial maps of $\Sz^{(c)}(\mathbf{r})$, (c,d \& e,f) time integrated spatial density distributions of the the spin-up and spin-down components of the condensed polariton system. The solid black bars in each panel represent $10~\mu$m.}
	\label{Fig3}
\end{figure}

Figure~\ref{Fig3} shows the simulated time-integrated spatial $\Sz^{(c)}(\mathbf{r})$ maps (a,b) along with the corresponding time integrated spin-up $|\psi_+|^2$ (c,d) and spin-down $|\psi_-|^2$ (e,f) condensate density distributions from simulations of cross-circularly polarized pumps with (b,d,f) and without (a,c,e) spin relaxation. When there is no spin relaxation, $\eta=0$, the two opposite circularly polarized pumps induce approximately a single potential peak for their respective spin resulting in two smooth radially expanding condensate envelopes (see Fig.~\ref{Fig3}(c) and~\ref{Fig3}(e)), not in agreement with experiment (compare Fig.~\ref{Fig3}(a) with Fig.~\ref{Fig2}(d)). When spin relaxation is included, $\eta \neq0$, each pump spot induces a potential peak for both spins (see Fig.~\ref{Fig2}(b)) resulting in strong scattering of the spins (see Fig.~\ref{Fig3}(d) and~\ref{Fig3}(f)), which yields a good reproduction of the features seen experimentally (see Fig.~\ref{Fig3}(b)).

\subsection{Gradually tuning the pumps polarization asymmetry}
\begin{figure}
	\centering
	\includegraphics[width=8.6cm]{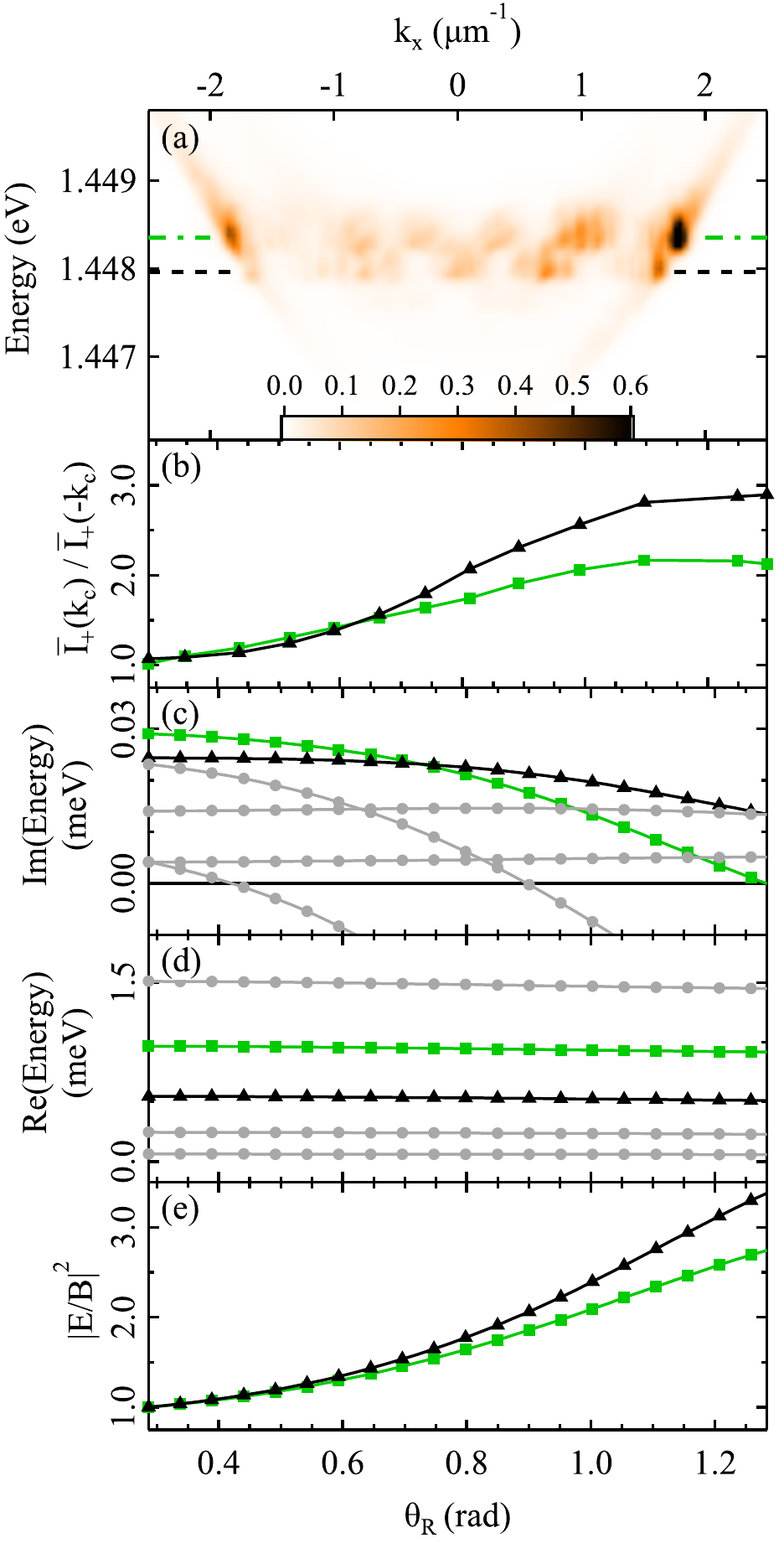}
	\caption{(a) Normalized dispersion of the RCP PL for $\theta_{L}\approx0.29$ and $\theta_{R}\approx1.30$ (i.e., $\SzL = - \SzR = 0.84$) plotted in a linear colorscale which has been saturated above 0.6 for clarity. (b) Experimentally measured population ratio of left $e^{- ik_c x}$ and right $e^{i k_c x}$ propagating spin-up polaritons across the whole dyad.  (c \& d) Imaginary and real parts respectively of the energies from Eq.~\eqref{Eqn_allowedEnergies} where the two highest imaginary valued energy branches (corresponding to condensate states with highest occupation) are shown by black triangles and green squares. (e) Analytical ratio ($|E/B|^2$) between the amplitudes of right and left propagating waves in the asymmetric barrier problem calculated via Eq.~\eqref{Eqn_LeftVsRightPropagation} for $\SzL = 0.84$. The dyad separation distance is approximately $10.5$ $\mu$m.} 
	\label{Fig4}
\end{figure}

By fixing the circular polarization of the left pump and varying the polarization of the right pump, direct control over the level of asymmetry in the potentials and thus the counter propagating cross-spin currents is achieved. This can be monitored by measuring the spectral density along the symmetry axis $k_y=0$ in reciprocal space (dispersion) of each spin component. In Fig.~\ref{Fig4}(a) we show the RCP PL dispersion for $\SzL=0.84\pm0.01$ and $\SzR=-0.84\pm0.01$ and a dyad separation distance of $d \approx 10.5\;\mu\mathrm{m}$. By gradually ramping $\SzR$ from $0.84\pm0.01$ to $-0.84\pm0.01$ we measure the change in the populated dispersion line profiles spectrally centered around the black dashed and green dot-dashed lines in Fig.~\ref{Fig4}(a). These two energy levels are simply the resonant modes of the double potential barrier system which were previously analyzed for the scalar polariton case~\cite{topfer_time-delay_2020}. We can then quantify the different amounts of left $e^{- ik_c x}$ and right $e^{ik_c x}$ propagating polaritons in the spin-up component in each line profile (i.e., at each energy). Here, the outflow wavevector $k_c$ corresponds to the free particle wavevector, which is energy-dependent with $k_c \approx 1.64 \mu\mathrm{m}^{-1}$ and $k_c \approx 1.79\mu\mathrm{m}^{-1}$ for the two modes highlighted in Fig.~\ref{Fig4}(a) by the black dashed and green dot-dashed lines respectively. For each mode we extract the left and right propagating polariton populations $\bar{I}_+(\pm k_c)$ by fitting Gaussian profiles to the spectral density peaks around $k_x =\pm k_c$ and integrating over. We point out that the same procedure can be applied to spin-down polaritons. Figure~\ref{Fig4}(b) shows the ratio, $\bar{I}_+(k_{c})/\bar{I}_+(-k_{c})$, evolving as $\SzR$ is changed from $0.84 \to -0.84$ for the two dominant energy states present in the dispersion. We observe that as $\SzR$ is rotated from being co- to cross-circularly polarized the ratio of the currents in both energy levels smoothly increases.

This observation can be qualitatively reproduced by considering a 1D non-Hermitian Schr\"{o}dinger equation describing the transmission and reflection properties of waves in an asymmetric barrier problem.
\begin{equation}
    i \hbar \frac{\partial \psi}{\partial t} = \left[- \frac{\hbar^2}{2m^*} \frac{\partial^2}{\partial x^2} + V(x) - \frac{i \hbar \gamma}{2} \right] \psi.
\end{equation}
We have dropped the spin index since the same analysis applies to both spinor components. Given the narrow waist of the laser spots we can approximate the double barrier potential with Dirac delta distributions,
\begin{equation}
    V(x) = V_L\delta(x+d/2) + V_R\delta(x-d/2).
\end{equation}
As described through Eqs.~\eqref{eq.pot0} and~\eqref{eq.res} the value of $V_{L,R}$ depends on the polarization $\SzL$ and $\SzR$ of their respective pump spots. The lasers DCP is defined,
\begin{equation}
    S_{z_{L(R)}} = \frac{P_{+_{L(R)}} - P_{-_{L(R)}}}{P_{+_{L(R)}} + P_{-_{L(R)}}},
\end{equation}
\begin{align}
P_{+_{L(R)}} & =  P_0 \cos^2{(\theta_{L(R)})} \\
P_{-_{L(R)}} & =  P_0 \sin^2{(\theta_{L(R)})},
\end{align}
where $\theta_{L(R)} \in [0, \pi/2]$. We can then express the left and right potentials for the spin-up polaritons as,
\begin{equation}
    V_{L(R)} = V_0 (\cos^2{( \theta_{L(R)} )} + \eta)
\end{equation}
where $V_0$ has absorbed the parameters of Eq.~\eqref{eq.pot0} for simplicity and, as before, $\eta$ is a free parameter describing the contribution coming from spin-relaxation within the laser generated exciton reservoirs.

The problem is decomposed into the three spatial regions,
\begin{align}
    \psi_{L}(x) &= Ae^{ikx}+Be^{-ikx}, \quad (x\leq -d/2), \\
    \psi_{M}(x) &= Ce^{ikx}+De^{-ikx}, \quad (-d/2<x\leq d/2), \\
    \psi_{R}(x) &= Ee^{ikx}+Fe^{-ikx}, \quad (x>d/2).
\end{align}
where $k$ lies in the first quadrant of the complex plane due to the non-Hermiticity of the problem. We can immediately set $A=0$ and $F=0$ as waves incident from the far left or right are not present. With the following simplification,
\begin{equation}
    v_{L,R} = \frac{2m^{*}V_{L,R}}{\hbar^{2}} 
\end{equation}
and applying the condition of wave function continuity at each delta potential we arrive at the following transcendental equation for $k$,
\begin{equation} \label{eq.transc}
1 = \frac{v_{L} v_{R} e^{2ikd}}{\left(2ik - v_{L}\right)\left(2ik - v_{R}\right)  }.
\end{equation}
The energy of the solutions is written,
\begin{equation}
    \epsilon(k) = \frac{\hbar^2 k^2}{2m^*} - \frac{i\hbar \gamma}{2}.
    \label{Eqn_allowedEnergies}
\end{equation}
The imaginary and real parts of the energies $\epsilon$ are plotted in Fig.~\ref{Fig4}(c) and~\ref{Fig4}(d) with gray circles and we have highlighted with black triangles and green squares the two solution branches which have the highest imaginary part for a given $\theta_R$ and a fixed $\SzL = 0.84$. High imaginary energy corresponds to solutions which the condensate dominantly occupies because these solutions have the largest gain. Therefore, the black and the green curves are the ones most relevant to the experimentally observed PL. An analytical expression for the ratio of right and left directed spin currents corresponding to the experimental results in Fig.~\ref{Fig4}(b) can be written as,
%
\begin{equation}
    \left| \frac{E}{B} \right|^2 =  \left| \frac{v_{L}}{2ik-v_{R}} \right|^2.
    \label{Eqn_LeftVsRightPropagation}
\end{equation}
Figure~\ref{Fig4}(e) shows the calculated values from Eq.~\eqref{Eqn_LeftVsRightPropagation} as the polarization of the right pump spot is rotated, qualitatively reproducing the behavior seen experimentally in Fig.~\ref{Fig4}(b). We note that for large $\theta_R$ the theory deviates from the experiment which could be due to both the finite width of the potential barriers and/or the two-dimensional nature of the system which allows polaritons to escape in the transverse direction ($y$-axis).

\section{Conclusions}
In summary, we have investigated the impact of non-resonant pump geometries with a non-uniform DCP on extended polariton condensate systems. The results underline the importance of spin relaxation mechanisms within the laser generated excitonic reservoirs which affects the interaction-induced potential landscape for each condensate spin component. We demonstrate the existence of optically controllable counter propagating spin currents in a spinor polariton dyad resulting in coherent matter-wave spin jets emitted from the system of two interacting condensates. The origin of these spin jets comes from the asymmetric potential landscape induced by the tunable asymmetric pump DCP. The ability to create a macroscopic polariton fluid with simple optical control over spin patterns and the ease of characterisation can potentially lead to new strategies to quantify the effects microscopic particle spin dynamics have on the fluid by changing the photon and exciton composition of the polariton quasiparticles.


\section{Acknowledgments}
The authors acknowledge the support of the UK’s Engineering and Physical Sciences Research Council (grant EP/M025330/1 on Hybrid Polaritonics), and the RFBR projects No. 20-52-12026 (jointly with DFG) and No. 20-02-00919. H.S. acknowledges hospitality provided by the University of Iceland. We thank M. Silva for their contribution to the experimental setup.

\appendix

\section{Simulations}\label{Methods_Simulation}

The observed features of the condensed systems are reproduced by numerically solving the driven-dissipative spinor 2D Gross-Pitaevskii equation coupled to an inactive and active exciton reservoirs~\cite{Carusotto_RMP2013} (see Eqs.~\eqref{Eqn_GPE}-\eqref{Eqn_GPE_InactiveRes}). The order parameter of the condensate spin components is given by $\psi_\pm$, $n_{A\pm}$ denotes the active reservoir density that resides in the dispersion bottleneck region and feeds the condensed state, and $n_{I\pm}$ denotes the inactive reservoir composed of high momentum excitons that feed into the active reservoir. 
\begin{widetext}
\begin{equation}
    i\frac{\partial \psi_{\pm}(\mathbf{r},t)}{\partial t} = \left(-\frac{\hbar\nabla^{2}}{2m^*} + \frac{i}{2}\left(Rn_{A_{\pm}}-\gamma\right) + \alpha_{1}|\psi_{\pm}(\mathbf{r},t)|^{2} +\alpha_{2}|\psi_{\mp}(\mathbf{r},t)|^{2} + g_{1}n_{A_{\pm}} + g_{2}n_{A_{\mp}}\right)\psi_{\pm}(\mathbf{r},t)
    \label{Eqn_GPE}
\end{equation}

\begin{equation}
    \frac{\partial n_{A_{\pm}}(\mathbf{r},t)}{\partial t} = -\left(\gamma_{A} + \gamma_{S} + R|\psi_{\pm}(\mathbf{r},t)|^2\right)n_{A_{\pm}}(\mathbf{r},t) + Wn_{I_{\pm}}(\mathbf{r},t)+\gamma_{S}n_{A_{\mp}}(\mathbf{r},t)
    \label{Eqn_GPE_ActiveRes}
\end{equation}

\begin{equation}
    \frac{\partial n_{I_{\pm}}(\mathbf{r},t)}{\partial t} = -\left(\gamma_{I} + \gamma_{S} + W\right)n_{I_{\pm}}(\mathbf{r},t) + P_{\pm}+\gamma_{S}n_{I_{\mp}}(\mathbf{r},t)
    \label{Eqn_GPE_InactiveRes}
\end{equation}
\end{widetext}
Here, $m^*$ is the polariton effective mass, $R$ is the spin-conserving stimulated scattering rate of polaritons from the active reservoir to the condensed state (we neglect stimulated spin-flip scattering from the reservoirs to the opposite spin condensate components), $\gamma$ is the polariton decay rate, $\alpha_{1}$ and $g_{1}$ are the parallel spin polariton-polariton and polariton-exciton interaction strengths and $\alpha_{2}$ and $g_{2}$ are the antiparallel spin polariton-polariton and polariton-exciton interaction strengths respectively. We adopt a fixed ratio between parallel and antiparallel spin interaction terms $\alpha_{2}=-0.2\alpha_{1}$ and $g_{2}=-0.2g_{1}$ similar to previous works~\cite{cerna_ultrafast_2013, vladimirova_polariton-polariton_2010}. $\gamma_{A,I}$ are the decay rates of active and inactive reservoir excitons respectively, $W$ is the conversion rate between inactive and active reservoir excitons, and $\gamma_S \propto \eta$ is the spin relaxation rate.

We note that there is no real or effective magnetic field, coupling the two spinor components, needed to explain the results of the experiment. Naturally, any cavity sample can possess some finite birefringence and/or TE-TM splitting which splits the in-plane polarized modes~\cite{Leyder_NatPhy2007} but in our experiment no obvious splitting was observed. However, due to the spin relaxation $\gamma_S$, the two spins depend indirectly on the density of each other through their active reservoirs $n_{A\pm}$. This is a purely non-linear effect that can stabilize the overall system but does not affect the relative phase between the spins which means that 
\begin{equation}
     \langle \psi_+^* \psi_- \rangle = 0,
\end{equation}
where $\langle . \rangle$ is an average over many random realizations of the condensate (stochastic initial conditions). We note that the value of $\gamma_{S}$ is set so that a fully circularly polarized pump, $\text{S}_{z_{L(R)}} = \pm 1$, yields $(n_{A+} - n_{A-})/(n_{A+}+n_{A-}) = \pm 0.11$, or i.e. 11\% spin polarized active reservoir, around condensation threshold where $|\psi_\pm|^2 \simeq0$. 

The parameters used in simulations are $m^*=0.28$meV/c$^{2}$, $\gamma=1/5.5$ ps$^{-1}$,  $\gamma_{A}=0.05$ ps$^{-1}$,  $\gamma_{I}=0.002$ ps$^{-1}$, $\gamma_{S}=0.05$ ps$^{-1}$, $\alpha_{1}=0.005$ ps$^{-1} \upmu$m$^{2}$, $\alpha_{2} = -0.2\alpha_{1}$,  $g_{1}=0.03$ ps$^{-1} \upmu$m$^{2}$, $g_{2} = -0.2g_{1}$, $W=0.05$ps$^{-1}$ and $R=0.107$ps$^{-1} \upmu$m$^{2}$. The pumps are Gaussian profiles with a full width half maximum of $2.2$ $\mu$m.


\bibliography{refs}

\end{document}